\documentclass[times,twocolumn]{aastex6}
\usepackage{bm}
\usepackage{graphicx}
\usepackage{lipsum}
\usepackage{mathtools}
\usepackage{hyperref}
\hypersetup{dvips, colorlinks=true, linkcolor=black, citecolor=black, filecolor=blue, urlcolor=black}

\newcommand{\Mbh}{M_\bullet}
\newcommand{\Jc}{J_\mathrm{c}}
\newcommand{\Jz}{J_z}
\newcommand{\Jlc}{J_\mathrm{lc}}
\newcommand{\DRR}{D_{JJ}^\mathrm{RR}}
\newcommand{\DNR}{D_{JJ}^\mathrm{NR}}
\newcommand{\dRR}{d_\mathrm{RR}}
\newcommand{\Tc}{T_\mathrm{c}}
\newcommand{\Nbody}{$N$-body\xspace}
\newcommand{\mstar}{M_{\star}}
\newcommand{\nur}{\nu_{\mathrm{r}}}
\newcommand{\num}{\nu_{\mathrm{M}}}
\newcommand{\rh}{r_\mathrm{h}}
\newcommand{\rt}{r_{\mathrm{t}}}
\newcommand{\nugr}{\nu_{\mathrm{GR}}}
\newcommand{\nup}{\nu_{\mathrm{p}}}
\newcommand{\nurr}{\nu_{\mathrm{r}}}
\newcommand{\rp}{r^{\prime}}
\newcommand{\np}{n^{\prime}}
\newcommand{\bI}{\mathbf{I}}
\newcommand{\bIp}{\mathbf{I}^{\prime}}
\newcommand{\br}{\mathbf{r}}
\newcommand{\bvel}{\mathbf{v}}
\newcommand{\thetap}{\theta^{\prime}}
\newcommand{\app}{a^{\prime}}
\newcommand{\Jp}{J^{\prime}}
\newcommand{\Jzp}{J_{z}^{\prime}}
\newcommand{\tp}{t^{\prime}}
\newcommand{\fpp}{f^{\prime}}
\newcommand{\deltaD}{\delta}
\newcommand{\wC}{\widehat{C}}
\newcommand{\Mtot}{M_\mathrm{tot}}

\newcommand{\rd}{\mathrm{d}}
\newcommand{\re}{\mathrm{e}}
\newcommand{\ri}{\mathrm{i}}
\newcommand{\rg}{r_{\mathrm{g}}}

\usepackage{xargs}
\newcommandx\dint[2][usedefault, addprefix=\global, 1=, 2=]{\!\!\int_{#1}^{#2}\!\! \rd}

\usepackage{soul}
\newcommand\eq{equation}

\newcommand\Eq{Equation}

\newcommand{\scrrpy}{{\small{\textsc{scRRpy}}}}

\makeatletter

\clearpage{}\usepackage{amsmath}
\usepackage[noabbrev]{cleveref}
\usepackage{placeins}
\usepackage{xspace}

\usepackage{acronym}
\newacro{sma}{semi-major axis}
\newacroplural{sma}[smas]{semi-major axis}
\newcommand{\sma}{\ac{sma}}
\newcommand{\smas}{\acp{sma}}
\newacro{MBH}{massive black hole}
\newacroplural{MBH}[MBHs]{massive black holes}
\newcommand{\MBH}{\ac{MBH}}

\newacro{BH}{black hole}
\newacroplural{BH}[BHs]{black holes}

\newacro{HMF}{Hamiltonian mean field}

\newacro{DC}{diffusion coefficient}
\newacroplural{DCs}{diffusion coefficients}
\newcommand{\DC}{diffusion coefficient}
\newcommand{\DCs}{diffusion coefficients}
\newacro{PDF}{probability distribution function}
\newacroplural{PDFs}{probability distribution functions}
\newcommand{\PDF}{\ac{PDF}}

\newacro{FP}{Fokker-Planck}

\newacro{FFT}{Fast Fourier Transform}

\newacro{DFT}{Discrete Fourier Transform}

\newacro{BL}{Balescu-Lenard}
\newcommand{\BL}{\ac{BL}}
\newacro{RR}{resonant relaxation}
\newcommand{\RR}{\ac{RR}}
\newacro{MC}{Monte-Carlo}

\newacro{BW}{Bahcall-Wolf}
\newcommand{\BW}{\ac{BW}}
\newacro{TDE}{tidal disruption event}
\newacroplural{TDE}[TDEs]{tidal disruption events}

\newcommand{\TDEs}{\acp{TDE}}
\newacro{EMRI}{extreme mass ratio inspiral}
\newacroplural{EMRI}[EMRIs]{extreme mass ratio inspirals}

\newcommand{\EMRIs}{\acp{EMRI}}
\newacro{GW}{gravitational wave}
\newacroplural{GW}[GWs]{gravitational waves}
\newcommand{\GW}{\ac{GW}}

\newacro{GR}{general relativity}
\newcommand{\GR}{\ac{GR}}

\usepackage{commath}

\clearpage{}
\makeatother

\begin{document}

\title{Scalar Resonant Relaxation of Stars Around A Massive Black Hole}

\author{Ben Bar-Or and Jean-Baptiste Fouvry\altaffilmark{1}}
\affil{Institute for Advanced Study, Princeton, NJ, USA}
\altaffiltext{1}{Hubble Fellow}


\begin{abstract}
In nuclear star clusters, the potential is governed by the central massive
black hole, so that stars move on nearly Keplerian orbits and the total
potential is almost stationary in time. Yet, the deviations of the potential
from the Keplerian one, due to the enclosed stellar mass and general
relativity, will cause the stellar orbits to precess. Moreover, as a result
of the finite number of stars, small deviations of the potential from
spherical symmetry induce residual torques that can change the stars' angular
momentum faster than the standard two-body relaxation. The combination of
these two effects drives a stochastic evolution of orbital angular momentum,
a process named ``resonant relaxation''. Owing to recent developments in the
description of the relaxation of self-gravitating systems, we can now fully
describe scalar resonant relaxation (relaxation of the magnitude of the
angular momentum) as a diffusion process. In this framework, the potential
fluctuations due to the complex orbital motion of the stars are described by
a random correlated noise, whose statistical properties are fully
characterized by the stars' mean field motion. On long timescales, the
cluster can be regarded as a diffusive system, whose diffusion coefficients
depend explicitly on the mean field stellar distribution through the
properties of the noise. We show here, for the first time, how the diffusion
coefficients of scalar resonant relaxation, for a spherically symmetric
system, can be fully calculated from first principles, without any free
parameters. We also provide an open source code that evaluates
these diffusion coefficients numerically.
\end{abstract}

\section{Introduction}

Nuclear star clusters with a central \MBH\ are dense environments where the
interactions between stars play a crucial role. Although they are among the
densest stellar environments in the Universe, their gravitational potential
is still dominated by the central \MBH\@. As a result, stars move on nearly
stationary Keplerian orbits. The gravitational potential from the stars
themselves only leads to small potential perturbations that modify the purely
Keplerian potential of the \MBH\@. Nevertheless, these small perturbations, as
well as the corrections from general relativity, are the ones that drive the
long-term evolution of the stellar cluster.

The evolution of a stellar system with a central \MBH\ is a classical problem
of stellar dynamics. It was first studied in the context of globular clusters
with a central \MBH\ by~\cite{Peebles1972}
and~\cite{Bahcall+1976,Bahcall+1977}. These seminal works showed, under
various simplifying approximations~\citep{Nelson+1999}, that the two-body
diffusion coefficients for a spherically symmetric and isotropic system can be
calculated from first principles, where the only unknown is the Coulomb
logarithm. This was subsequently generalized by~\citet{Shapiro+1978}
and~\citet{Cohn+1978}, who derived a two-dimensional diffusion equation (in
energy and angular momentum) and calculated the associated diffusion
coefficients~(see, e.g.,~\cite{Bar-Or+2016}). Although the existence of
central black holes in globular clusters is still unknown, many nuclear star
clusters contain a massive black hole in their center~(see~\cite{Graham2016}
for a review).

In addition to the standard two-body relaxation driven by local scatterings,
there exists in galactic nuclei a more efficient mechanism to change the
angular momentum of stars. This process, named \RR\ by~\citet{Rauch+1996},
results from the coherent motion of the stars along their nearly fixed
Keplerian orbits: a given test star will be subject to residual torques
persisting on long timescales.
\RR\ can be separated into two different processes, scalar \RR\
that drives the evolution of the eccentricity, i.e.\ the magnitude of the angular
momentum, and vector \RR\ that drives the orbital orientation. The
residual torques associated with scalar \RR\ are randomized by the in-plane
orbital precession. The residual torques associated with vector \RR\ persist on
longer timescales, as they are randomized by the changes of the
orbital orientations themselves. This implies that the orbital evolution by vector
\RR\ is much faster than the one by scalar \RR\, but can only affect the
direction of the angular momentum vector~\citep{Rauch+1996,
Hopman+2006a}. Extensive studies of vector \RR\ were presented by~\citet{Kocsis+2011,
Kocsis+2015, Roupas+2017}. Here, our main focus is scalar \RR\@.

Scalar \RR\@, which can dominate the angular momentum's evolution, over the
standard two-body relaxation, did not have a formal self-consistent description
for many years. Previous attempts at modeling this process were only
qualitative, and many studies had to use ad-hoc methods to include
it~\citep[e.g.,][]{Hopman+2006a, Madigan+2011, Merritt+2011, Antonini+2013,
Merritt2015a}. Recent advances in $N$-body simulations allowed for the study
of \RR\ numerically~\citep{Merritt+2011, Hamers+2014}, but were lacking a fully
self-consistent theory.

Only recently, several studies~\citep{Bar-Or+2014, Sridhar+2016,Fouvry+2017a},
put forward independently the foundation for a self-consistent kinetic theory
of \RR\@. Building upon these works, we show here that in the case of an
isotropic spherical system, scalar RR can be described as a diffusion process,
for which one can derive and calculate the \DCs\ from first principles.

This paper is organized as follows. In Section~\ref{s:hamiltonian} we write
the orbit-averaged Hamiltonian of a test star, as a Fourier sum over the
orbital angles of both the test star and the field stars. Following the
$\eta$-formalism~\citep{Bar-Or+2014,Fouvry+2017b}, these Fourier components are
the random noise terms that drive the stochastic evolution of the test star's
orbital angular momentum, $J$. In Section~\ref{s:dc} we follow this approach
to write a closed expression for the diffusion coefficient of
scalar \RR\@. In Section~\ref{s:dcs} we briefly discuss the two-dimensional
($a$, ${J/\Jc}$) structure of the \DCs\ and compare it to two-body
relaxation. Finally, we summarize our results and discuss future applications
in Section~\ref{s:summary}.

\section{Hamiltonian}
\label{s:hamiltonian}

Let us consider a star of mass $\mstar$ moving on a nearly Keplerian orbit of
\sma\ $a$ around a \MBH\ of mass $\Mbh$ embedded in a spherically symmetric
and isotropic in velocity star cluster of density ${\rho(r)}$.
The orbits of the stars in the
potential of the \MBH\ can be described in angle-action variables. In this
case, it is convenient to use the Delaunay
variables~\citep{BinneyTremaine2008}, where the three actions are:
${\Jc=\sqrt{G\Mbh a}}$, the maximal (circular) angular momentum for a given
$a$, ${J=\sqrt{1-e^2}\Jc}$, the specific orbital angular momentum with $e$
the eccentricity, and ${\Jz=J\cos(\theta)}$, the $z$ component of the
angular momentum with $\theta$ the inclination angle w.r.t.\ an inertial
reference frame. The corresponding angles are: the mean anomaly $\mathcal{M}$,
the argument of pericenter $\omega$, and the longitude of the ascending node
$\Omega$.

Following~\cite{Bar-Or+2014}, we use the addition theorem for spherical
harmonics to write the secular (orbit-averaged) specific Hamiltonian of the test
star as a multipole expansion
\begin{equation}
\label{eq:H}H=H_0(a,J)+\!\!\!\!\sum_{m,n=-\infty}^{\infty}\!\!\!\!\re^{\ri(m\Omega+\ri n\omega)}\eta_{nm}(a,J,\Jz,t),
\end{equation}
In this equation, the first term is the mean field potential, while the second
term describes the potential fluctuations around the mean field due to the
intricate motion of the finite number of field stars.

The mean field potential reads
\begin{equation}
H_{0}(a,J)=\Phi_\mathrm{MBH}(a)+\Phi_\mathrm{GR}(a,J)+\overline{\Phi}_\star(a,J).\label{eq:H0}
\end{equation}
It is composed of the Keplerian potential of the central \MBH,
\begin{equation}
\label{eq:phi_mbh}\Phi_\mathrm{MBH}(a)=-\frac{1}{2}\,\nur(a)\Jc,
\end{equation}
where ${\nurr(a)=\sqrt{G\Mbh/a^{3}}}$ is the fast orbital frequency
imposed by the central \MBH\@, an effective correction to the
Keplerian potential,
\begin{equation}
\label{eq:phi_gr}\Phi_\mathrm{GR}(a,J)=-3\frac{\rg}{a}\frac{\Jc}{J}\nurr(a)\Jc,
\end{equation}
which reproduces the orbit-averaged Schwarzschild (in-plane) orbital precession,
where ${\rg=G\Mbh/c^2}$, and ${\overline{\Phi}_\star(a,J)}$ the mean
field potential due to the stellar cluster around the \MBH\@.

The last two terms of $H_0$ induce respectively a prograde and retrograde
in-plane orbital precession, and the combined precession is
\begin{equation}
\label{eq:nu_p}\dot{\omega}\equiv\nup(a,J)=\dpd{H_0}{J}=\nugr(a,J)+\num(a,J),
\end{equation}
where
\begin{equation}
\label{eq:nugr}\nugr(a,J)=\dpd{\Phi_\mathrm{GR}}{J}=3\frac{\rg}{a}\frac{\Jc^2}{J^2}\nurr(a),
\end{equation}
is the precession induced by \GR\@, and
\begin{equation}
\label{eq:num}\num(a,J)=\dpd{\overline{\Phi}_\star}{J}=\frac{\nur(a)}{\pi M_\bullet e}\dint[0][\pi]f\Mtot(r[f])\cos f
\end{equation}
is the mass induced precession~\citep{Kocsis+2015}, where ${\Mtot(r)}$ is the
stellar mass enclosed within a radius $r$, and $f$ is the true anomaly.

The last term in \eq~\eqref{eq:H} is due to the discrete nature of the stellar potential
and describes the fluctuations around the mean field potential due to the
motion of the field stars. Note that this is the only time-dependent term in
the Hamiltonian and therefore this term drives the orbital diffusion of the test
star. These terms take the form
\begin{equation}
\label{eq:eta}\eta_{nm}(a,J,\Jz,t)=\!\!\sum_{k=1}^N G M_k\!\!\!\!\sum_{\np=-\infty}^{\infty}\!\!\!\!\re^{-\ri(m\Omega_k(t)+\np\omega_k(t))}\psi_{m n\np}(\bI,\bI_{k}(t)),
\end{equation}
where the first sum is over the $N$ field stars and $M_k$ is the mass of the
$k$-th field star. In the large $N$ limit, ${\eta_{nm}}$ can be considered as
random Gaussian noise terms, with ${\langle\eta_{nm}\rangle=0}$. In
\eq~\eqref{eq:eta}, we also
introduced the angular Fourier components of the pairwise orbit-averaged
interaction potential given by
\begin{equation}
\label{eq:psi}\psi_{m n\np}(\bI,\bIp)=-\!\!\!\!\sum_{\ell=\ell_{\min}}^{\infty}\!\!\!G^{\ell}_{m n\np}(\theta,\thetap)\,K_{n\np}^{\ell}(a,J,\app,\Jp),
\end{equation}
with
\begin{align}
\label{eq:Kl}K_{n\np}^{\ell}(a,J,\app,\Jp)={}&\left\langle K_{\ell}(r,\rp)\,\re^{\ri(n f-\np\fpp)}\right\rangle_{\!\!\circlearrowright}\!\!\nonumber\\={}&\left\langle K_{\ell}(r,\rp)\cos(nf)\cos(\np\fpp)\right\rangle_{\!\!\circlearrowright},
\end{align}
where ${\ell_{\min}\equiv\max\{1,|m|,|n|,|\np|\}}$, ${\bI=(\Jc,J,\Jz)}$ stands for the action vector,
${\langle\,\cdot\,\rangle_\circlearrowright}$ denotes the orbit-average,
and ${K_{\ell}(r,\rp)={\min(r,\rp)}^{\ell}/{\max(r,\rp)}^{\ell+1}}$ is
the usual min-max term from the Legendre expansion of the Keplerian potential.
Here, we note that the component ${\ell=0}$ does not contribute to the diffusion.
Finally, ${G^{\ell}_{m n\np}(\theta,\thetap)}$ is the geometrical factor
\begin{equation}
\label{eq:G}G^{\ell}_{m n\np}(\theta,\thetap)=\frac{4\pi\,y_{\ell}^{n}\,y_{\ell}^{\np*}}{2\ell+1}d_{nm}^{\ell}(\theta)d_{\np m}^{\ell}(\thetap),
\end{equation}
where
${d_{n m}^{\ell}(\theta)}$, related to
the Wigner's rotation matrices ${D_{nm}^\ell(\alpha,\beta,\gamma)=\re^{-\ri n\alpha}d_{n m}^\ell(\beta)\re^{-\ri m\gamma}}$~\citep[e.g.,][]{Rose1995}, satisfies
${{\langle|d_{n m}^{\ell}|}^2\rangle_{\theta}=2/(2\ell\!+\!1)}$, and
${y_{\ell}^{n}=Y_{\ell}^{n}(\pi/2,\pi/2)}$ satisfies
${4\pi{|y_{\ell}^{n}|}^2/(2\ell\!+\!1)=[(k_+\!-\!1)!!(k_{-}\!-\!1)!!]/[(k_{+})!!(k_{-})!!]}$ and is non zero only if
${k_\pm=\ell\pm n}$ are even.

Thus, this geometrical factor satisfies
\begin{equation}
\label{eq:G_symmetry}{\langle{|G^{\ell}_{m n\np}(\theta,\thetap)|}^2\rangle}_{\theta,\thetap}=\frac{16\pi^2{|y_{\ell}^{n}|}^2{|y_{\ell}^{\np}|}^2}{{(2\ell+1)}^{4}},
\end{equation}
where
${\langle\,\cdot\,\rangle_{\theta,\thetap}=\!\int\!\rd\theta\rd\thetap\sin(\theta)\sin(\thetap)\,(\cdot)}$ is the average over the
inclination angles of the field and test stars.

\section{The diffusion coefficients}
\label{s:dc}

In this section we connect the stochastic Hamiltonian
in \eq~\eqref{eq:H},
which describes the motion of a test particle for a given set realization of
the field stars, to the diffusion equation describing the evolution of the
angular momentum of test particles undergoing stochastic
perturbations induced by the field stars from the stellar cluster.

Here, we assume a spherically symmetric stellar distribution for the cluster where the
phase-space density of stars ${f(\br,\bvel)=f(E,J)}$
depends only on the orbital (positively defined) energy ${E=G\Mbh/2a}$ and
$J$. The number of stars per unit $a$, per unit $J$ is given by
${N(a,J)=4\pi^{3}(2J/\Jc)f(E,J)G\Mbh}$. It is convenient to write
${N(a,J)=N(a)f_J(J;a)}$, where ${N(a)}$ is the number of stars per unit
$a$ and ${f_J(J;a)}$ is the \PDF\ of $J$ for a given $a$. In the simplifying
case where ${f(E,J)\propto{|E|}^p}$ (with ${p<3/2}$), one has
${N(a)=(3-\gamma)\tfrac{N_{0}}{a}(\tfrac{a}{a_{0}})^{2-\gamma}}$,
where ${\gamma=p\!+\!\tfrac{3}{2}}$ and
${N_0=g(\gamma)N(<\!a_0)}$, with
${g(\gamma)=\tfrac{\sqrt{\pi}}{2^\gamma}\tfrac{\Gamma(1+\gamma)}{\Gamma(\gamma-1/2)}}$
and ${N(<\!a_0)}$ is the number of stars within a radius $a_0$.

The relaxations in energy and angular momentum are usually treated as two
separate one-dimensional relaxation processes, where the system first relaxes in
angular momentum with fixed energy, and then relaxes in
energy. In the absence of a loss-cone, during the first stage the system relaxes to an isotropic
angular momentum distribution ${f_J(J;a)=2J/\Jc^2}$. This is a general
result of maximal entropy and is independent of the exact details of the
relaxation process, i.e.\ different relaxation processes can change the timescale on
which the system relaxes but not the steady state. This steady state is
slightly modified by the existence of a loss-cone, where stars with
${J<\Jlc(a)}$ are lost, e.g.\ by tidal disruption with
${\Jlc\simeq\sqrt{2\rt G\Mbh}}$, where
${\rt={(\Mbh/\mstar)}^{1/3}R_\star}$ is the tidal radius. For compact objects
and stars with tidal disruption radius smaller than $8 G\Mbh/c^2$, the
loss-cone is given by ${\Jlc\simeq 4G\Mbh/c}$, for which orbits plunge
directly into the \MBH\@. Finally, the existence of a loss-cone logarithmically suppresses the distribution of
angular momentum toward $\Jlc$, so that
${f_J(J;a)\propto(2J/\Jc)\log(J/\Jc)}$~\citep[e.g.,][]{Bar-Or+2016}.

Following~\citet{Bar-Or+2014} and~\citet{Fouvry+2017b}, the \PDF\ of $J$, at a
given \sma\ $a$, $P(J,t;a)$, evolves according to a diffusion (Fokker-Plank)
equation of the form
\begin{equation}
\label{eq:FP}\dpd{P(J,t;a)}{t}=\frac{1}{2}\dpd{}{J}\left[J\DRR(a,J)\dpd{}{J}\frac{P(J,t;a)}{J}\right],
\end{equation}
where the diffusion coefficient, ${\DRR(a,J)}$, is proportional to the power spectrum of the
noise terms $\eta_{nm}$ evaluated at the precession frequency ${\nup(a,J)}$, so that
\begin{equation}
\label{eq:DRR_I}\DRR(a,J)=2\sum_{n=1}^{\infty}n^2\wC_{n}(a,J,n\nup(a,J)),
\end{equation}
where ${\wC_{n}(J,n\nup(a,J))}$ is the Fourier transform,
${\hat{f}(\omega)=\!\int_{-\infty}^{\infty}\!\rd t f(t)\re^{\ri\omega t}}$, of
the correlation function
\begin{equation}
\label{eq:Cft}C_{n}(a,J,t-\tp)=\sum_{\mathclap{m=-\infty}}^{\infty}\;\;\!\!\int_{-J}^{J}\!\!\frac{\rd\Jz}{2J}\langle\eta_{nm}(a,J,\Jz,t)\eta_{nm}^*(a,J,J_z,\tp)\rangle,
\end{equation}
and we used the fact that ${\wC_{n}(a,J,n\nup(a,J))}$ is invariant under
${n\to-n}$, to sum only over positive $n$, which introduces a factor $2$ w.r.t.~\citet{Fouvry+2017b}.

The correlation function in \eq~\eqref{eq:Cft} depends on time through the
motion of the field stars. As the Keplerian orbits of the field stars
evolve, the cluster's potential changes, and on long timescales, the system is
reshuffled and the potential fluctuations become uncorrelated. Here, the main
source of orbital evolution is the apsidal precession of the orbits due to the
enclosed stellar mass, $\Mtot(r)$, as well as the relativistic in-plane
precessions. Assuming that the field stars are moving on Keplerian orbits
precessing in-plane because of the mean field Hamiltonian $H_0$
(\eq~\eqref{eq:H0}) (and ignoring collective effects~\citep[see][]{Fouvry+2017b}), the diffusion
coefficient from \eq~\eqref{eq:DRR_I} can be written explicitly as
\begin{align}
\label{eq:DRR_II}\DRR(a,J)\!=&4\pi G^2\sum_{i}M_i^2\sum_{n=1}^{\infty}\!\sum_{\np=-\infty}^{\infty}\!\!\!\!n^2\!\dint\app N_i(\app)\!\dint\Jp f_{J,i}(\Jp;a^\prime)\nonumber\\&\hspace{-1.8cm}\times{|A_{n\np}(a,J,\app,\Jp)|}^2\,\deltaD(n\nup(a,J)-\np\nup(\app,\Jp)),
\end{align}
where we considered a mass spectrum ${\{M_i\}}$ of field stars and defined a susceptibility
coefficient which is averaged over both $\Jz$ of test star and the field star
\begin{align}
\label{eq:A}&{|A_{n\np}(a,J,\app,\Jp)|}^2=\!\!\!\!\sum_{\ell=\ell_{\min}}^\infty\!\sum_{m=-\ell}^{\ell}\!\int_{-J}^{J}\!\!\!\frac{\rd\Jz}{2 J}\!\!\int_{-\Jp}^{\Jp}\!\!\!\frac{\rd\Jzp}{2\Jp}\,{|\psi_{m n\np}(\bI,\bIp)|}^{2}\nonumber\\&\!\!\!=\!\!\!\sum_{\ell=\ell_{\min}}^\infty\!\!\!\frac{16\pi^2{|y_{\ell}^{n}|}^2{|y_{\ell}^{\np}|}^2}{{(2\ell+1)}^{3}}\,{|K_{n\np}^{\ell}(a,J,\app,\Jp)|}^{2}.
\end{align}
By solving the resonant condition in \eq~\eqref{eq:DRR_II}, we can carry out
the integral over $\Jp$ to obtain
\begin{align}
\label{eq:DRR_III}\DRR(a,J)={}&4\pi G^2\sum_i M_i^2\!\!\sum_{n,\np=1}^{\infty}\!\!\frac{n^2}{\np}\dint\app N_i(\app)\nonumber\\&\times\sum_{\Jp}\frac{f_{J,i}(\Jp;\app){|A_{n\np}(a,J\!,\app\!,\Jp)|}^2}{|\partial_{\Jp}\nup(\app\!,\Jp)|},
\end{align}
where the sum on $\Jp$ runs over the solutions, $J_{+}$ and $J_{-}$, of the
resonant conditions ${\nup(\app,J_\pm)=\pm(n/\np)\,\nup(a,J)}$,
which depend on $\app$, $a$, $J$ and on the ratio, ${n/\np}$, of the
resonance numbers.

The information about the underlying cluster is contained in the angular
momentum distribution function ${f_{J,i}(J;a)}$, in the mass weighted \sma\
distribution ${\sum_i M_i^2 N_i(a)}$, and in the stellar contribution to the
precession which enters the resonant condition ${\deltaD(n\nup(a,J)-\np\nup(\app,\Jp))}$ in \eq~\eqref{eq:DRR_II}, while scaling with mass as ${\sim\sum_i(M_i/\Mbh)N_i(<a)\nur(a)}$ (see
\eq~\eqref{eq:num}). In a multi-mass population, the system will
undergo a strong mass segregation, where heavier masses will develop a steeper
density slope than the lighter ones~\citep{Alexander+2009}. This means that
at small \smas\ the heavy stars will dominate the diffusion, which in turn increases
the diffusion rate by the heavy to light mass ratio. For simplicity, in the
upcoming applications, we limit ourselves to a single-mass
population.

\Eq~\eqref{eq:DRR_III} is the main result of this work. It shows that for a
spherically symmetric and isotropic stellar distribution the \DCs\ associated
with \RR\ can be derived and calculated from first principles. Carrying out the
integral in \eq~\eqref{eq:DRR_III} is conceptually straightforward but can be
technically challenging. It requires solving the resonant condition and
integrating over $\app$ and over the two true anomalies $f$ and $\fpp$ in
${|A_{n\np}|^{2}}$ (see \eq~\eqref{eq:Kl}). We provide a code,
\scrrpy\footnote{Available at \url{https://github.com/benbaror/scrrpy}.}, in which the integration
is carried out using the Vegas Monte-Carlo integration scheme~\citep{Lepage1978}.

The Fokker-Planck equation~\eqref{eq:FP} can be rewritten in the more
traditional form
\begin{equation}
\label{eq:FP_II}\dpd{P(J,t)}{t}=-\dpd{}{J}D_{J}P(J,t)+\frac{1}{2}\dpd{}{{J^2}}D_{JJ}P(J,t),
\end{equation}
where the two-diffusion coefficients satisfy the fluctuation-dissipation
relation
\begin{equation}
D_J=\frac{1}{2 J}\,\frac{\partial(J D_{JJ})}{\partial J},\label{fluctuation_dissipation_relation}
\end{equation}
with ${D_{JJ}=\DRR}$.

Assuming an isotropic $J$ distribution and a single-mass population,
\eq~\eqref{eq:DRR_III} can be written as
\begin{equation}
\label{eq:DRR_SM}\DRR(a,J)=2\tilde{\tau}^2\Tc(a)\dRR(a,J),
\end{equation}
where ${\tilde{\tau}=(\mstar/\Mbh)\sqrt{N(<2 a)}\Jc\nurr(a)}$ is the
typical strength of the residual torque~\citep{Gurkan+2007},
${\Tc(a)\sim 1/\num(a)}$ is the typical coherence time (see below), ${N(<2a)}$ is the number of
stars with \smas\ smaller than ${2a}$, and
\begin{align}
\label{eq:dRR_SM}\dRR(a,J)={}&\!\!\sum_{n,\np=1}^{\infty}\!\!\frac{n^2}{\np}\dint\app\frac{N(\app)}{N(<2a)}\nonumber\\&\times\sum_{\Jp}\frac{4\pi\Jp a^2{|A_{n\np}(a,J\!,\app\!,\Jp)|}^2}{{\Jc(\app)}^2|\partial_{\Jp}\nup(\app\!,\Jp)|\Tc(a)},
\end{align}
is dimensionless.

\begin{figure}[htbp!]
\begin{center}
\includegraphics[width=0.48\textwidth]{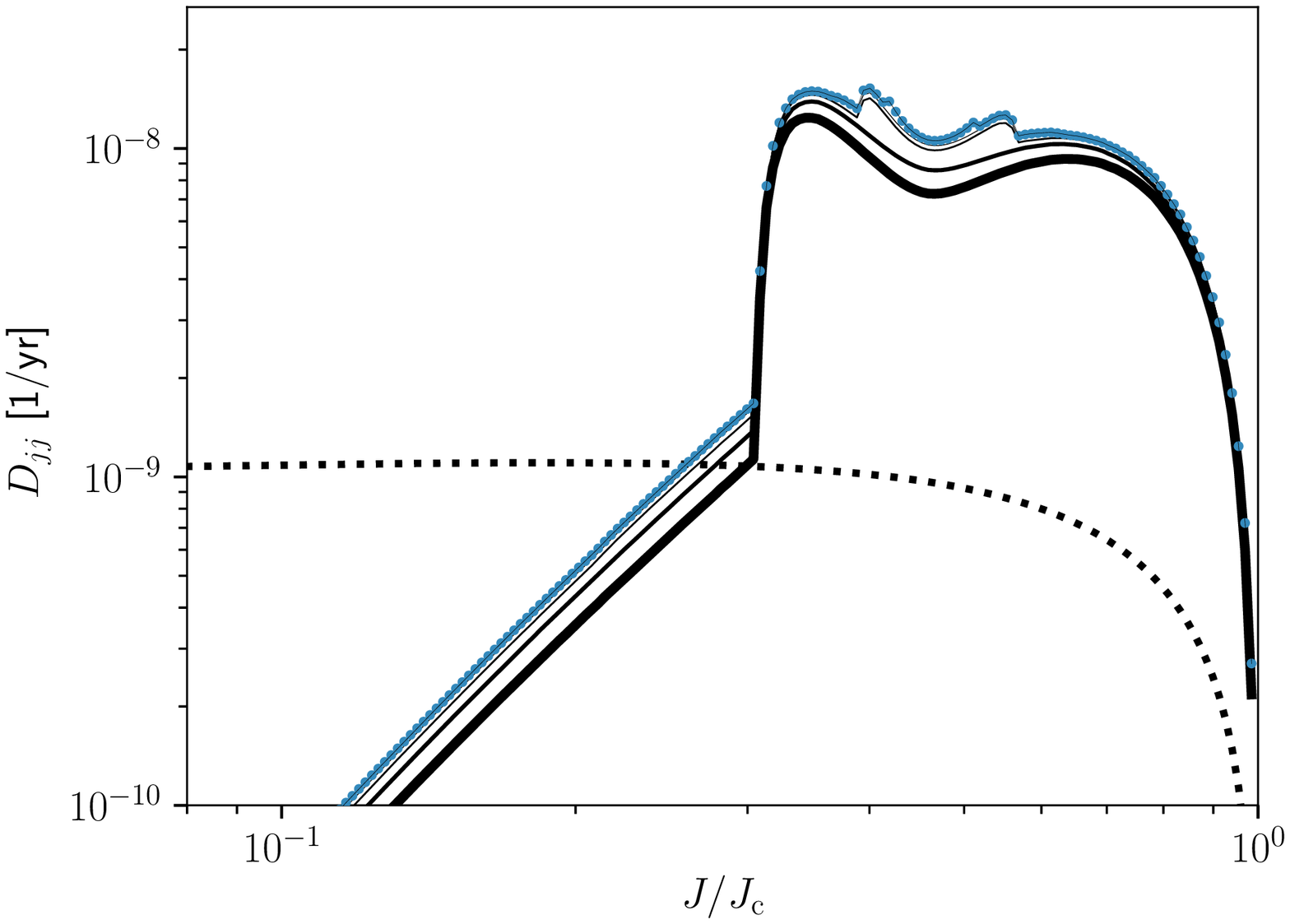}
\caption{\label{fig:DRR}~Normalized angular momentum ${j=J/\Jc}$
diffusion coefficients as a function of $j$ for two-body (dotted line)
and resonant relaxation (\RR, dots). The \RR\ diffusion coefficient is
calculated using~\eq~\eqref{eq:DRR_III}, while the two-body relaxation
time is calculated from \eq~(125) in~\cite{Bar-Or+2016}. The sharp drop
in the \RR\ diffusion coefficient occurs where the precession frequency
of the test star ${\nup(j)}$ is comparable to the coherence frequency
of the system ${1/\Tc(a)}$ (see Figure~\ref{fig:drr}). We also show
the convergence w.r.t.\ the maximal harmonic number $\ell$ in black
lines, from ${\ell=1}$ (thickest line) to ${\ell=5}$ (thinnest
line with dots). In this figure, the diffusion coefficients are evaluated
at ${a\simeq 8\,\mathrm{mpc}}$ for an isotropic \BW\ cusp
${f(E)\propto E^{1/4}}$ of solar mass stars around a \MBH\ of
${M_\bullet={4\times 10^6 M_\odot}}$ and a total stellar mass $M_\bullet$
within the radius of influence ${\rh=2\mathrm{pc}}$, using
\scrrpy\@.}
\end{center}
\end{figure}

Since we assumed that the cluster is isotropic in velocities, i.e.\ ${\partial f(J,E)/\partial J=0}$, there is no
dynamical friction\footnote{Here we define dynamical friction as the drift term
proportional to the mass of the subject star, as opposed to the parametric
drift proportional to the mass of the field stars~\citep{Chavanis2012}.}, and no
amplification through collective effects~\citep{Fouvry+2017b}. In the absence
of a loss-cone, the zero flux steady state solution reads
${f(J)=2J/\Jc^{2}}$. In practice, ${f(J)}$ is logarithmically suppressed near
the loss-cone, $\Jlc$, and therefore deviates from the isotropic ${f(J)\propto J}$
distribution. As a result, both dynamical friction and collective effects can
become important near the loss-cone. However, as \RR\ is quenched near the loss
cone (see Figure~\ref{fig:DRR}), these effects are expected to be of no
practical importance for the overall relaxation (which also includes two-body
relaxation).

In Figure~\ref{fig:DRR} we show the \RR\ \DC\ for the normalized angular
momentum ${j=J/\Jc}$, with the notation ${D_{jj}=\DRR/\Jc^{2}}$, as a function of $j$
and compare it with two-body
relaxation~\citep[][\eq~{(125)}]{Bar-Or+2016} at
${a\simeq 8\,\mathrm{mpc}}$. Here, we assume a stellar population with a \BW\ cusp
density profile ${\rho(r)\propto r^{-7/4}}$ and stars of one solar mass
${M=M_\odot}$ around a \MBH\ of mass ${\Mbh=4\times 10^{6}M_{\odot}}$, with
a total stellar mass $M_\bullet$ within the radius of influence
${\rh=2\,\mathrm{pc}}$. The various bumps appearing in Figure~\ref{fig:DRR}
are associated with the respective contributions from different resonance pairings ${(n,\np)}$.
Figure~\ref{fig:DRR} also demonstrates the fast
convergence of $\DRR$ w.r.t\ the harmonic number $\ell$. Unlike two-body
relaxation which has a logarithmic divergence (manifested in the Coulomb
logarithm), \RR\ has no divergences.

\begin{figure}[htbp!]
\begin{center}
\includegraphics[width=0.48\textwidth]{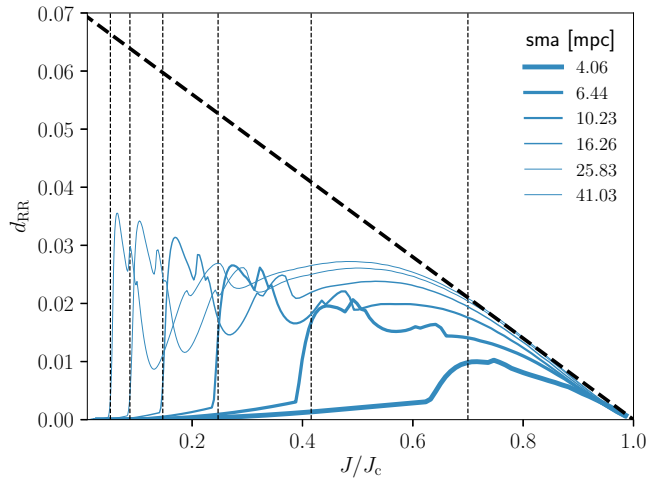}
\caption{\label{fig:drr}~The dimensionless diffusion coefficient $\dRR$ (\eq~\eqref{eq:dRR_SM})
as a function of $j$ at different \smas\@. At small eccentricity (${1-j\ll 1}$), $\dRR$ can be
approximated by ${\dRR\approx\langle\tau^2\rangle/\tilde{\tau}^2\approx 0.07(1-j)}$ (dashed line). When the relativistic precession approaches ${\sim 0.45/\Tc}$, $\dRR$ drops as $j$ becomes an adiabatic invariant.
The cusp's parameters are the same as in Figure~\ref{fig:DRR}.
}
\end{center}
\end{figure}

As shown in Figure~\ref{fig:drr}, near ${J=\Jc}$ (i.e.\ ${e\ll 1}$),
$\DRR$ is well approximated by ${\DRR\simeq 2\Tc\langle\tau^2\rangle}$,
where ${\langle\tau^2\rangle/\tilde{\tau}^2\simeq 0.07(1-j)}$ is the
averaged residual torque~\citep{Bar-Or+2016}, and $\Tc$ is the typical
correlation time of the system. Here, we also find that the ansatz
${\Tc\simeq\sqrt{\pi/2}\,\nup^{-1}(2a,1/\sqrt{2})}$
from~\cite{Bar-Or+2016}, for which the correlation time is proportional to the
median precession time evaluated at ${2a}$ (the median eccentricity is
${e=j=1/\sqrt{2}}$), provides a good approximation
of $\DRR$, as long as ${\nup(2a,1/\sqrt{2})}$ is dominated by mass
precession, i.e.\ most of the field stars at this \sma\ are non-relativistic.

This implies that for non-relativistic orbits, \RR\ scales as
${\DRR/\Jc^2\sim(M/\Mbh)\nur(a)}$ which is independent of the number of
stars~\citep{Rauch+1996, Hopman+2006a},
while two-body relaxation scales like
$\DNR\sim N(<a)(M^2/M_\bullet^2)\nur(a)\log\Lambda$. Therefore, since
$N(<a)\le M/M_\bullet$, \RR\ can be significantly faster that two-body
relaxation in some regions of orbital space~\citep[e.g.,][]{Bar-Or+2016}.

When the precession frequency of the test star, ${\nup=\num+\nur}$, approaches
${1/\Tc}$, $\DRR$ sharply drops as it enters the relativistic regime where
the precession frequency of the test star is higher than the precession of the
bulk of the field stars, and $J$ becomes an adiabatic invariant. In Figure~\ref{fig:drr}
we show that this suppression of \RR\ occurs at $j_0$ where
${\nugr(a,j_0)\simeq 0.45/\Tc}$.

\section{Discussion}
\label{s:dcs}

In this section, we briefly investigate the phase-space structure of $\DRR$,
compare it to the standard two-body relaxation \DCs\@, and comment about its
contribution to various physical phenomena in the vicinity of the \MBH\@. We
use \scrrpy\ to calculate $\DRR$
and for simplicity, we consider as previously a stellar cluster composed of a
single-mass population with a \BW\ power-law density cusp
${\rho(r)\propto r^{-7/4}}$.

\begin{figure*}[htbp!]
\begin{center}
\includegraphics[width=0.9\textwidth]{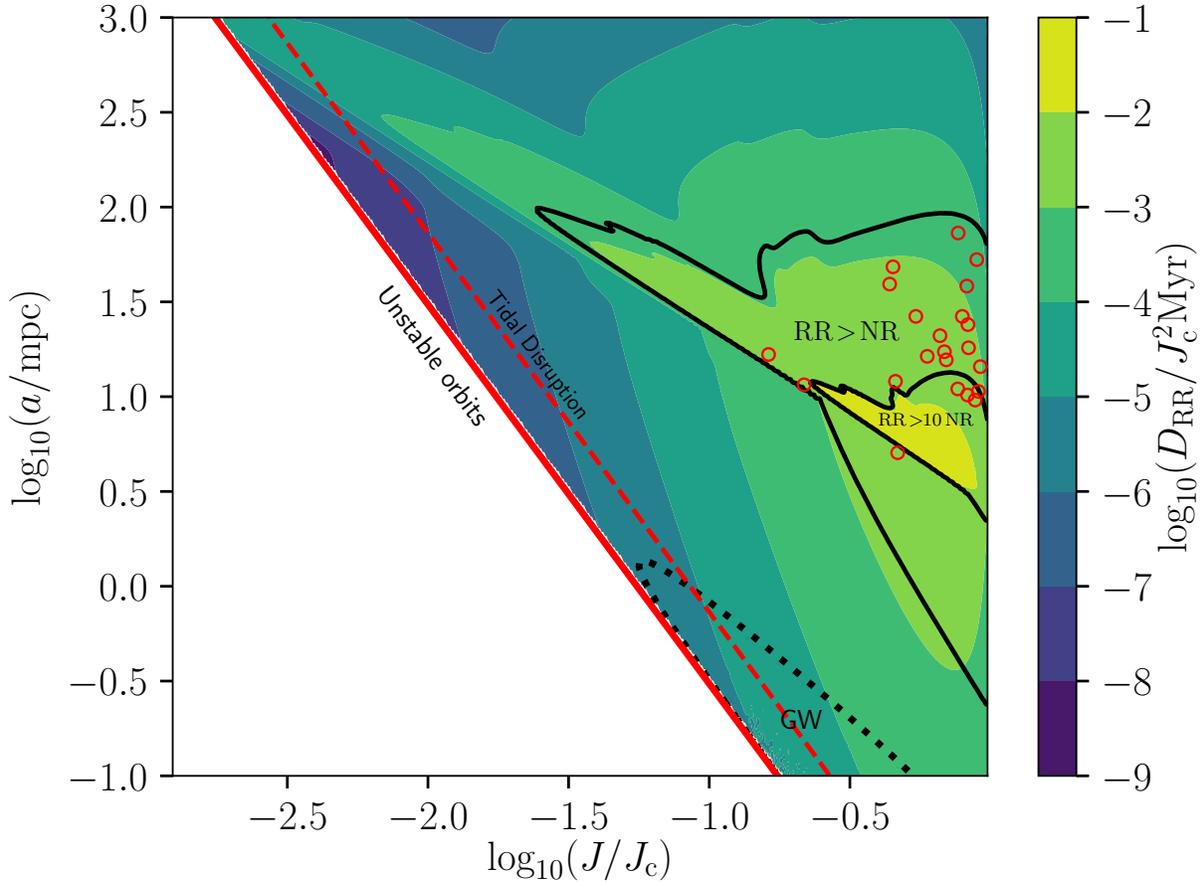}
\caption{\label{fig:phasespace}~Phase space structure of
${\DRR/\Jc^2}$, showed as a color map on a logarithmic scale. Diffusion
by \RR\ is faster than two-body or non-resonant relaxation (NR) in a limited region of
phase-space (solid black contours), which is far from the region where
gravitational wave emission dominates the orbital evolution (dashed contour). The orbits of
S-stars observed in the Galactic center~\citep{Gillessen+2017} (red
circles), lie in the \RR-dominated region. Orbits beyond the relativistic
loss-cone, ${\Jlc=4G\Mbh/c}$ (solid red line), are short-lived. Solar
mass stars will tidally disrupt if their orbital pericenter distance is
smaller than
${\rt={(\Mbh/M_\odot)}^{1/3}R_\odot}$~\citep{Alexander2017} (dashed
red line). The cusp's parameters are the same as in Figure~\ref{fig:DRR}.
}
\end{center}
\end{figure*}

As shown in Figure~\ref{fig:phasespace}, diffusion by \RR\ can be faster than
two-body relaxation in a limited region of phase space. Interestingly, the
orbits of the young stellar population cluster in the Milky-Way Galactic
center~\citep[the S-stars cluster,][]{Ghez+2003, Schödel+2003,
Gillessen+2009,Gillessen+2017} are within this region.

At low ${j=J/\Jc}$ and low $a$, \RR\ is quenched by adiabatic invariance. This
is because the relativistic precession $\nugr$ increases as ${1/(a j^2)}$
(see \eq~\eqref{eq:nu_p}), and when ${\nugr(j)}$ is larger than the coherence
frequency $1/\Tc$, the \DC\ decays rapidly, as demonstrated in
Figure~\ref{fig:DRR}. In Figure~\ref{fig:phasespace}, this translates to a
line in ${(a,J/\Jc)}$ phase space where RR is quenched and two-body
relaxation takes over. This line is associated with the so-called
``Schwarzschild barrier'' that is observed in \Nbody\
simulations~\citep{Merritt+2011}. At large \sma, the mass precession time becomes
comparable to the orbital time and two-body relaxation wins over \RR\@.

Generally, event rates associated with loss-cone dynamics like \TDEs\ and
binary disruptions will be governed either by the dynamics near the boundary
between full- and empty loss-cone or by the dynamics near the radius of
influence $\rh$, depending on which radius is smaller. Typically, for a \MBH\ with
a mass of ${10^{5}\!-\!10^{7}\,M_\odot}$, these are of the same order of
magnitude~\citep[e.g.,][]{Alexander+2017}. Near the radius of influence, the
precession time is comparable to the orbital period and thus the dynamics is
governed by two-body relaxation. Therefore, \RR\ is not expected to have a
significant effect on these rates.

As shown in Figure~\ref{fig:phasespace}, \RR\ can have an effect on the
dynamics of stars deep in the cusp. \citet{Hopman+2006a} suggested that \RR\
can significantly increase the rates of \EMRIs\@. Additionally, \Nbody\
simulations~\citep{Merritt+2011, Brem+2014}, which inherently contain both \RR\
and two-body relaxation, showed that in practice the rates of \EMRIs\ are
comparable to the ones obtained by considering only two-body relaxation.
\citet{Bar-Or+2016} showed that as long as the region where the \RR\ diffusion
dominates over two-body relaxation is far from the region where \GW\ emission
dominates the orbital evolution (see dotted contour in
Figure~\ref{fig:phasespace}), \RR\ will not contribute significantly to the
\EMRIs\ event rate. This is indeed the case for the cusp considered in
Figure~\ref{fig:phasespace}.

Both \RR\ and two-body relaxation will drive the stellar distribution toward an
isotropic distribution in angular momentum, i.e.\ ${f(J)=2J/\Jc^{2}}$, or
$f(e)=2e$ in eccentricity, when neglecting loss-cone effects. This will
happen over the relaxation timescale, which is of order of the diffusion
timescale ${T_j(a)=1/D_{jj}^\mathrm{iso}(a)}$, where
${D_{jj}^\mathrm{iso}(a)=\!\int\!\rd j\,2 j D_{jj}(a,j)}$ is the
isotropic averaged \DC\@.

\begin{figure}[htbp!]
\begin{center}
\includegraphics[width=0.48\textwidth]{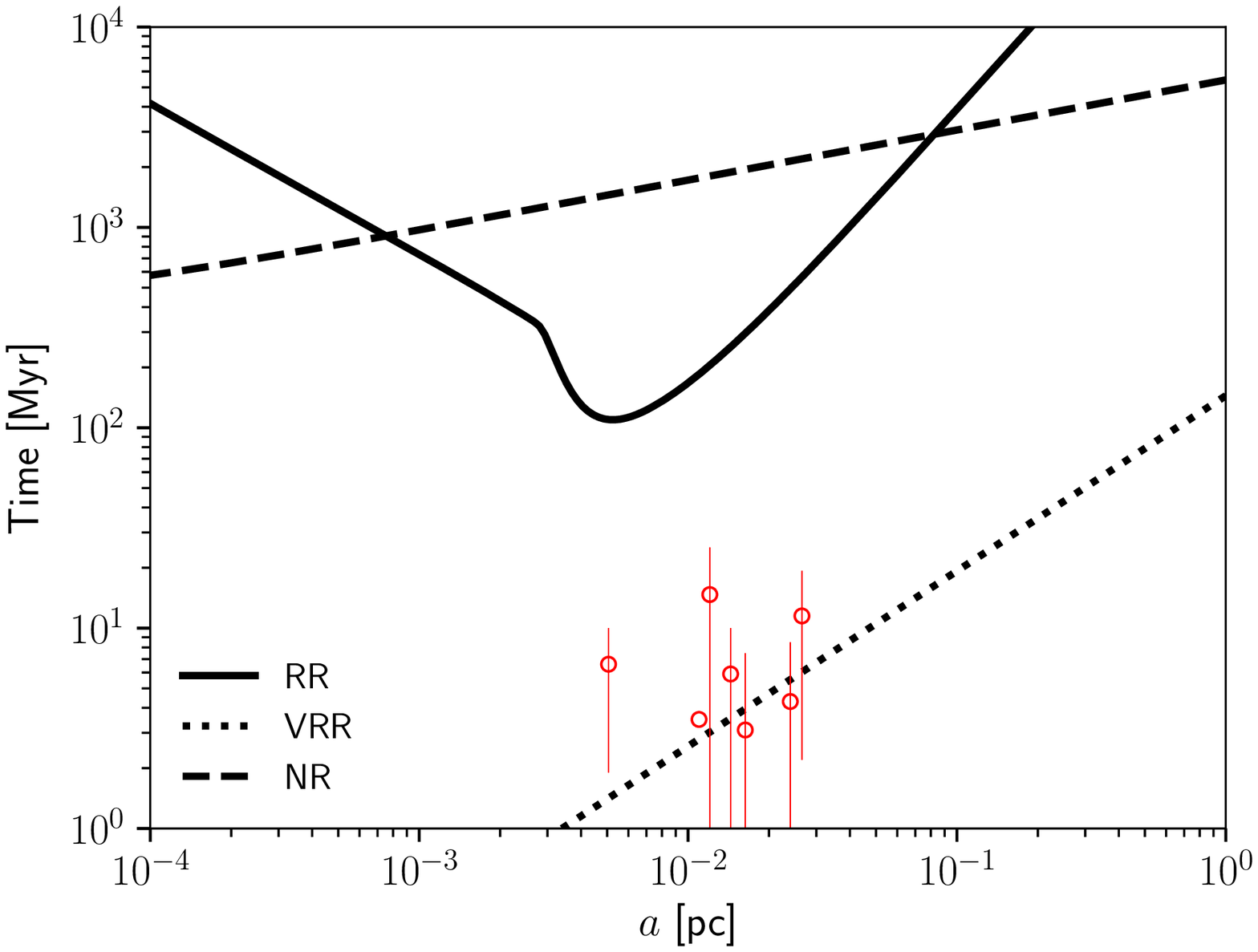}
\caption{\label{fig:timescales}~Angular momentum relaxation timescales
${T_j=1/D_{jj}^\mathrm{iso}}$ by \RR\ (solid line), and by
two-body relaxation (NR) (dashed line), as a function of semi-major
axis. Relaxation by \RR\ is faster than two-body relaxation
in a limited range of semi-major axis. For comparison, we show the main
sequence ages of some of the S-stars estimated recently
by~\cite{Habibi+2017} (red circles with error bars) and the vector \RR\
timescale (dotted line).
The cusp's parameters are the same as in Figure~\ref{fig:DRR}.
}
\end{center}
\end{figure}

In Figure~\ref{fig:timescales} we show the relaxation times for \RR\ and for
two-body relaxation. While the two-body diffusion time scales as
${T_{j}^\mathrm{NR}(a)\sim{(\Mbh/M_\star)}^2 P(a)/(N(<a)\log\Lambda)}$,
where ${P(a)}$ is the orbital period, the \RR\ diffusion time scales as
${T_{j}^\mathrm{RR}(a)\sim(\Mbh/M_\star)P(a)}$ in the region where the
precession is dominated by mass precession. As shown in
Figure~\ref{fig:timescales}, \RR\ can be significantly faster than two-body
relaxation deep in the cusp until it is quenched by the relativistic
precession. Although faster than two-body relaxation, the \RR\ diffusion
timescale is longer than the ages of some of the young stars observed in our
Galactic center~\citep{Habibi+2017}. This suggests that these stars did not have
the time to relax by \RR\ to the current nearly thermal ${f(e)\propto e}$
distribution observed today~\citep{Gillessen+2017}. Let us note however that in
a multi-mass system, one has
${T_{j}^\mathrm{RR}(a)\sim\Mbh\langle\mstar\rangle/\langle{\mstar}^2\rangle P(a)}$ and the diffusion time can be shorter by an order of
magnitude in regions where stellar black holes dominate the total enclosed
mass, as expected from strong mass segregation~\citep{Alexander+2009}.

An additional relaxation mechanism that will randomize the orbital orientation
is the so-called ``vector resonant relaxation''~\citep{Rauch+1996,
Hopman+2006a, Kocsis+2015}. As shown in Figure~\ref{fig:timescales} this
process can randomize the orientations of the orbits (but not their
eccentricities) on a shorter timescale
${T_{j}^{\mathrm{VRR}}\simeq(\Mbh/M_\star)(P(a)/\sqrt{N(<a)})}$~\citep{Kocsis+2015}.

\section{Summary}
\label{s:summary}

Relaxation processes in dense stellar systems around a \MBH\ are a classical
problem of stellar dynamics. Understanding these processes
is crucial for the long-term steady-state stellar distribution of nuclear clusters and mass
segregation therein, short-term transient phenomena such as tidal
disruptions, gravitational wave emissions, and hypervelocity stars and the
distribution of unique source populations such as young stars, X-ray binaries
and radio pulsars to name a few~\citep{Alexander2017}.

All these phenomena depend both on the relaxation in energy and in angular
momentum. The relaxation in energy is well described by two-body relaxation,
where the \DCs\ can be calculated from first principles for an isotropic
distribution function. The only poorly determined quantity is the Coulomb
logarithm, which has only a small effect on the diffusion.

Despite the approximation made in the derivation of these
\DCs~\citep[e.g.,][]{Nelson+1999}, they are in a good agreement with the ones
measured in direct \Nbody\ simulations~\citep{Bar-Or+2013}. However, relaxation
in angular momentum can be dominated by \RR~\citep{Rauch+1996,
Hopman+2006a}. While this was demonstrated by~\citet{Eilon+2009} and especially
by~\citet{Merritt+2011} using direct \Nbody\ simulations, a complete and
self-consistent theory of \RR\ was still lacking. The foundation for a concrete
kinetic theory of \RR\ was put forward independently
by~\citet{Bar-Or+2014},~\cite{Sridhar+2016}, and~\cite{Fouvry+2017a}.
In~\citet{Fouvry+2017b}, we generalized the method of~\citet{Bar-Or+2014} to a
general stochastic Hamiltonian with integrable mean field and showed it to be
equivalent to the (degenerate) \BL\ and Landau equations~\citep{Heyvaerts2010,
Chavanis2012, Chavanis2013}. This means that the different approaches
of~\citet{Bar-Or+2014, Sridhar+2016} and~\citet{Fouvry+2017a}, although
different in details, are essentially equivalent.

Building upon~\citet{Bar-Or+2014} and~\cite{Fouvry+2017b}, we presented here,
for the first time, a calculation of the scalar \RR\ \DCs\ from first
principles and without any free parameters. This brings to a closure the long
journey, started by~\cite{Rauch+1996}, of bringing the kinetic theory of \RR\
to the same level of completeness as the standard two-body relaxation
one. Although this treatment is limited to the diffusion of the angular
momentum magnitude in a spherical and isotropic background distribution, for
which collective effects can be ignored~\citep[e.g.,][]{Nelson+1999}, the same
limitations also apply to standard two-body relaxation
(see~\cite{Vasiliev2015} for applications to non-spherically symmetric systems).
Here, we also assumed that the \MBH\ dominates the potential. This assumption
will break down close to the radius of influence, where the contribution of the
underlying stellar population is comparable to that of the \MBH\@. This is not
a significant limitation as \RR\ is negligible compared to two-body relaxation
at this point (see Figure~\ref{fig:timescales}). Some of these limitations
could be mitigated in following studies, and the entire kinetic theory could be
tested against future \Nbody\ simulations that are already approaching a
realistic number of stars in galactic nuclei~\citep{Panamarev+2018}.

The ability to calculate \RR\ \DCs\ provides us with the opportunity to make
more realistic estimates on the effects of \RR\ on astrophysical phenomena in
galactic nuclei. As shown in Figure~\ref{fig:timescales}, \RR\ can dramatically
reduce the relaxation time in angular momentum. As a result, even short-lived
populations (like the young S-star cluster) can be relaxed to a thermal
eccentricity distribution. As \RR\ can efficiently drive the angular momentum
evolution, it may contribute to the supply rate of stellar objects into the
loss-cone. This contribution is significant only if the loss-cone is close to
the region where \RR\ dominates the diffusion over two-body relaxation and will
depend on the underlying stellar distribution and the specific loss-cone
scenario. We show that for a standard stellar population following a \BW\ cusp
around a ${4\times 10^6 M_\odot}$ \MBH, the contribution of \RR\ to the
\EMRIs\ and \TDEs\ rates is negligible.

\section*{Acknowledgements}
We are grateful to Tal Alexander, Adrian Hamers, Kirill Lezhnin, John
Magorrian, Christophe Pichon and Scott Tremaine for fruitful discussions. BB
acknowledges support from the Schmidt Fellowship. JBF acknowledges support
from Program number HST-HF2--51374 which was provided by NASA through a grant
from the Space Telescope Science Institute, which is operated by the
Association of Universities for Research in Astronomy, Incorporated, under NASA
contract NAS5--26555. This research is carried out in part within the
framework of the Spin(e) collaboration (ANR-13-BS05-0005,
\url{http://cosmicorigin.org}).

\bibliographystyle{apj}

\end{document}